\documentclass{aa}
\usepackage{graphicx}
\usepackage{txfonts}
\usepackage{natbib}
\bibpunct{(}{)}{;}{a}{}{,}
\usepackage{longtable}
\usepackage{lscape}
\usepackage{times}
\usepackage{textcomp}

\begin{document}

\title{Radio properties of the optically identified supernova remnant G107.0+9.0
\thanks {Reduced images (FITS) shown in Figs.~4-9 and Fig.~11 are available 
at the CDS via anonymous ftp to ...}}
%
\author{Wolfgang~Reich\inst{1}, Xuyang~Gao\inst{2,3,4} and Patricia~Reich\inst{1}}
\titlerunning{Radio properties of G107.0+9.0}
\authorrunning{W. Reich et al.}

\offprints{wreich@mpifr.de}

\institute{
  Max-Planck-Institut f\"{u}r Radioastronomie, Auf dem H\"{u}gel 69,
  53121 Bonn, Germany; \\{\it wreich@mpifr-bonn.mpg.de, preich@mpifr-bonn.mpg.de} \and National Astronomical Observatories, Chinese Academy of
  Sciences, Jia-20 Datun Road, Chaoyang District, Beijing 100101, 
  China;\\{\it xygao@nao.cas.cn}\and School of Astronomy and Space Sciences, University of Chinese Academy of Sciences, Beijing 100049, China \and
  CAS Key Laboratory of FAST, National Astronomical Observatories, Chinese Academy of Sciences, Beijing 100101, China}

\date{Received; accepted}

\abstract
{The vast majority of Galactic supernova remnants (SNRs) were detected by their synchrotron radio emission. 
Recently, the evolved SNR G107.0+9.0 with a diameter of about $3\degr$ or 
75~pc up to 100~pc in size was optically detected with an indication of faint associated radio emission. This SNR requires a detailed radio study. } 
{We aim to search for radio emission from SNR G107.0+9.0 by analysing new data from the Effelsberg 100-m and the Urumqi 25-m radio telescopes  
in addition to available radio surveys.}  
{Radio SNRs outside of the Galactic plane, where confusion is rare, must be very faint if they have not 
been identified so far. Guided by the H$\alpha$ emission of G107.0+9.0, we separated its radio emission from the  Galactic 
large-scale emission.}
{Radio emission from SNR G107.0+9.0 is detected between 22~MHz and 4.8~GHz with a steep non-thermal spectrum, 
which confirms G107.0+9.0 as an SNR. Its surface brightness is among the lowest known for Galactic SNRs. Polarised emission is 
clearly detected at 1.4~GHz but is fainter at 4.8~GHz. We interpret the polarised emission as being caused by a Faraday screen 
associated with G107.0+9.0 and its surroundings. Its ordered magnetic field along
the line of sight is below 1~$\mu$G. At 4.8~GHz, we identified a depolarised filament along the western periphery of G107.0+9.0 with
a magnetic field strength along the line of sight $B{_{||}} \sim 15~\mu$G, 
which requires magnetic field compression.}
{G107.0+9.0 adds to the currently small number of known, evolved, large-diameter, low-surface-brightness Galactic SNRs. 
We have shown that such objects can be successfully extracted from radio-continuum surveys despite the dominating large-scale
diffuse Galactic emission. }


\maketitle

\section{Introduction}

When searching for faint planetary nebula, \citet{Yuan13} noticed H$\alpha$ emission from an almost spherical object with a diameter of about
3$\degr$, which they proposed to be a so far unidentified supernova remnant (SNR).
\citet{Fesen20} confirmed G107.0+9.0  
as an SNR based on optical imaging in several lines and spectroscopic observations, which revealed shock velocities between 70~km s$^{-1}$  
and 100~km s$^{-1}$.  \citet{Fesen20} estimated the distance of G107.0+9.0 between 1.5~kpc and 2~kpc, implying a size of 75~pc to 100~pc. The SNR 
is likely in the radiative phase, expanding in a low-density interstellar medium about 235~pc to 315~pc above the Galactic
plane. 

Optically identified SNRs are rare, while most SNRs are identified by their non-thermal radio emission. 
The radio emission of new optically detected SNRs must be unusually faint or suffer from confusion with unrelated emission in the Galactic plane.
Otherwise, they would already have been identified from the numerous existing all-sky or Galactic plane surveys in radio continuum. 
However, G107.0+9.0 is located well outside of the strong emission along the Galactic plane, where confusion with other Galactic objects at 
radio wavelength is low in general. Diffuse large-scale Galactic emission, however, dominates at low frequencies, so objects like G107.0+9.0
with a diameter of about $3\degr$  are either masked or cause just small intensity fluctuations.

\citet{Fesen20} noticed a coincidence with faint radio-continuum emission in the medium-resolution 
1420-MHz northern-sky survey by \citet{Reich82}. This indicates that radio emission should be visible in low-frequency surveys as well,
and we checked the 22-MHz survey \citep{Roger99} and the 408-MHz survey \citep{Haslam82}. The area of G107.0+9.0
is also covered by the high-latitude extension of the Canadian Galactic plane survey (CGPS in the following) at 408~MHz \citep{Tung17} and 1420~MHz \citep{Landecker10}.
These maps have arc-minute angular resolution and are corrected for missing large-scale emission by single-dish data.
So far unpublished data from the Effelsberg Medium Latitude Survey (EMLS) at 1.4~GHz, which will comprise the area of the northern Galactic plane 
up to $\pm20\degr$, and a high-latitude extension of the Sino-German 
4.8-GHz survey of the Galactic plane include the area of G107.0+9.0. These surveys have an angular resolution of $9\farcm4$ and $9\farcm5$ and
include linear polarisation.

In Sect.~2, we describe the H$\alpha$ and radio data used in this study and present maps of G107.0+9.0 and
its surrounding area. In Sect.~3, we discuss the results of flux integration of G107.0+9.0 to derive its spectrum. We
analyse the observed linear polarisation in the direction of the SNR and apply a Faraday-screen model. 
Section~4 summarises our results. 

\section{H$\alpha$ and radio data used}

\subsection{H$\alpha$ emission}

\begin{figure}
\centering
\includegraphics[angle=-90, width=0.50\textwidth]{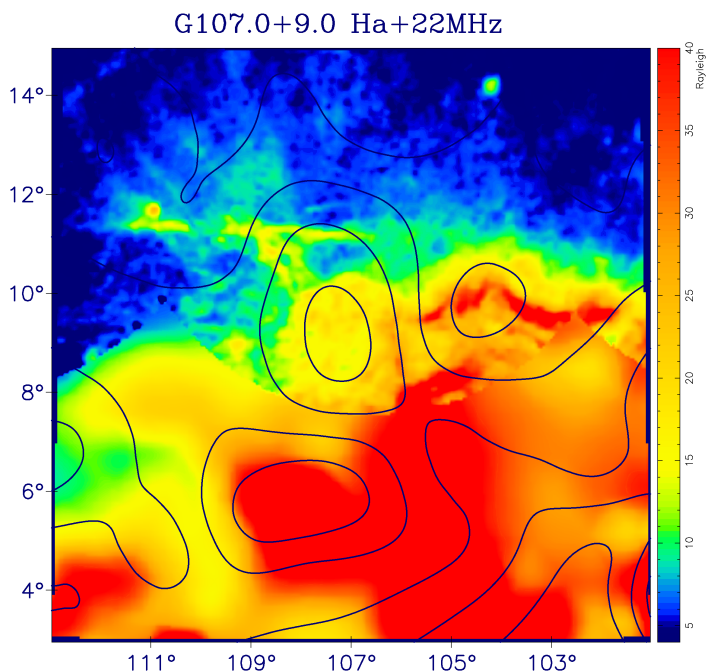}
\caption{Colour-coded H$\alpha$ map centred on G107.0+9.0 with overlaid 22-MHz contours from Fig.~\ref{22MHz}
to indicate coincident radio emission from G107.0+9.0. } 
\label{Ha}
\end{figure}

\citet{Fesen20} presented high-quality optical images and spectra to identify G107.0+9.0 as an SNR.   
Here, we used the public all-sky H$\alpha$ map with 6$\arcmin$ angular resolution or less by \citet{Finkbeiner03}, 
which is a combination of various H$\alpha$ surveys and includes G107.0+9.0. Details are given by \citet{Finkbeiner03}.
We extracted the H$\alpha$ map of G107.0+9.0 and its surroundings (Fig.~\ref{Ha}), 
which shows faint, almost circular H$\alpha$ emission from G107.0+9.0. Below
a Galactic latitude of 8$\degr$, the H$\alpha$ data have a resolution lower than 6$\arcmin$. 
We superimposed 22-MHz contours of G107.0+9.0 on the H$\alpha$ map (Fig.~\ref{Ha}) for clarity, because outside of G107.0+9.0,
more intense, unrelated H$\alpha$ emission is visible, which indicates the presence of thermal gas.
The background-subtracted H$\alpha$ intensity in the area of G107.0+9.0 is estimated to be between 3~Rayleigh and 7~Rayleigh, while
a western filament along the periphery of G107.0+9.0 (also seen in Fig.~\ref{CGPS408} and Fig.~\ref{CGPS+EMLS}) has up to about 12~Rayleigh.

\subsection{22-MHz data}

\begin{figure}
\centering
\includegraphics[angle=-90, width=0.50\textwidth]{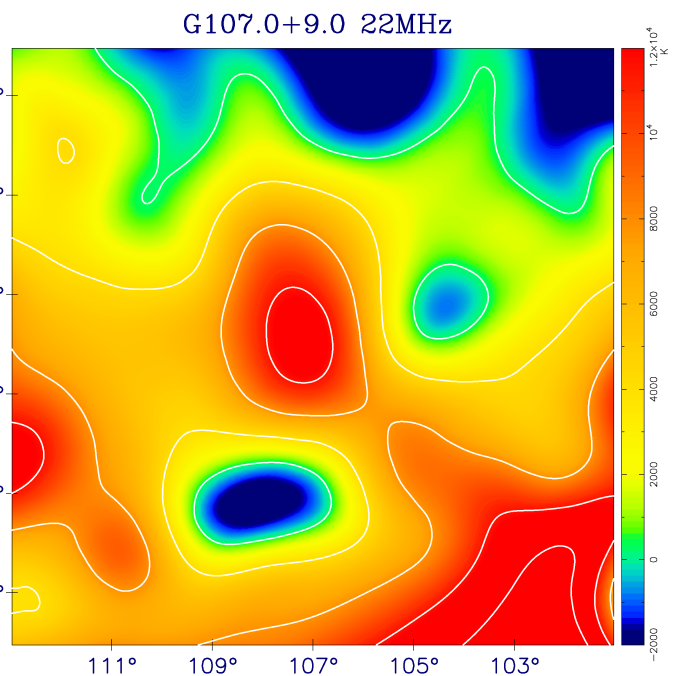}
\caption{Colour-coded total-intensity 22-MHz map of G107.0+9.0 after subtraction 
of large-scale emission exceeding  $5\degr$ by `unsharp masking' (see text). 
The superimposed 22-MHz contours are 4000~K apart.}
\label{22MHz}
\end{figure}

We extracted the area of G107.0+9.0 from the 22-MHz northern-sky survey conducted with the DRAO interferometer  \citep{Roger99}.
The angular resolution is EW - NS $1\fdg1 \times (1\fdg7$ $\times$ secant zenith angle), which means that the 22-MHz emission from
G107.0+9.0 with about $3\degr$ in diameter is barely resolved.
We applied the  `unsharp-masking' or `background-filtering' method \citep{Sofue79} to separate G107.0+9.0 
emission from the dominating large-scale Galactic emission, and we show the 22-MHz map without structures exceeding $5\degr$ in extent
in Fig.~\ref{22MHz}. Centrally peaked emission at the quoted SNR position is clearly visible.
We fitted the emission from G107.0+9.0 by a two-dimensional Gaussian, where a twisted plane defines the local zero-level,  
and we calculated a flux density of 468 $\pm$ 60~Jy,
which is based on the 22-MHz flux density of the nearby SNR Tycho with 680 $\pm$ 54~Jy as quoted by \citet{Roger86}.
We searched for strong discrete sources in the area of G107.0+9.0, which might have an influence on
the flux density of G107.0+9.0. The source 4C +66.24 is in the field and appears  
to be excessive compared to all other compact sources listed in the area of G107.0+9.0
and its surroundings. \citet{Vollmer10} used published flux densities of 4C +66.24 between 38~MHz and 4.85~GHz and
fitted a spectrum. Its extrapolation gives a flux density at 22~MHz of about 10~Jy, 
which reduces the flux density of G107.0+9.0 to 458 $\pm$ 60~Jy.

\subsection{408-MHz data}

\begin{figure}
\centering
\includegraphics[angle=-90, width=0.50\textwidth]{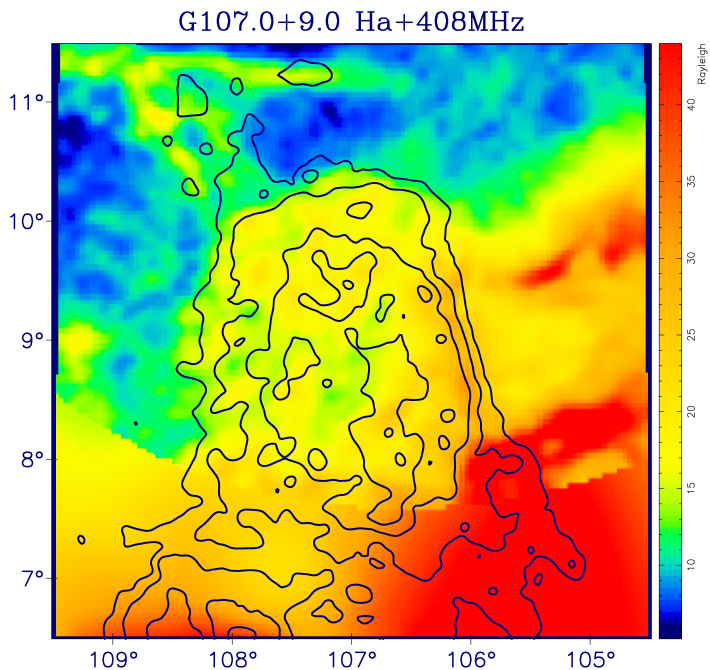}
\caption{Colour-coded H$\alpha$ map centred on G107.0+9.0 with overlaid 408-MHz contours from the processed CGPS 
survey \citep{Tung17}. The contours are 2~K $T_\mathrm{b}$ apart.}
\label{CGPS408}
\end{figure}

G107.0+9.0 is barely visible in the 51$\arcmin$-resolution 408-MHz all-sky survey maps \citep{Haslam82}. It is much better
traced by the CGPS high-latitude extension at 408 MHz \citep{Tung17}, which 
combines single-dish, large-scale emission from the 408-MHz all-sky survey \citep{Haslam82} and interferometric data. 
The CGPS 408-MHz survey has an angular resolution of 2$\farcm8 \times 2\farcm8~\times$ cosec(dec).
We processed the CGPS 408-MHz map by removing small-scale structures, like 
unresolved sources, and also large-scale emission. We separated compact sources and 
extended emission using the background-filtering method \citep{Sofue79} with a filtering beam of 5$\arcmin$, which results 
in a map with small-scale structures up to 5$\arcmin$ and a larger-scale intensity map, where the sum of both maps gives the original map.
The larger scale map was processed in the same way, but now removing structures exceeding about 3$\degr$.
We noted a non-uniform emission gradient towards higher Galactic 
latitudes, which makes it difficult to trace the outer boundary of the emission from G107.0+9.0. We could partly remove 
this emission gradient by defining new zero-levels for each row of the map by setting the edge areas to zero. We 
show the result of our processing in Fig.~\ref{CGPS408},
where 408-MHz contours are superimposed on the H$\alpha$-emission map (Fig.~\ref{Ha}). The 408-MHz emission coincides well with  
the H$\alpha$ emission from G107.0+9.0.
We integrated the filtered 408-MHz emission of G107.0+9.0 up to a radius of $1\fdg5$ and obtained 
a flux density of 23~$\pm$ 3~Jy.

\subsection{1.4-GHz data}

\begin{figure}
\centering
\includegraphics[angle=-90, width=0.50\textwidth]{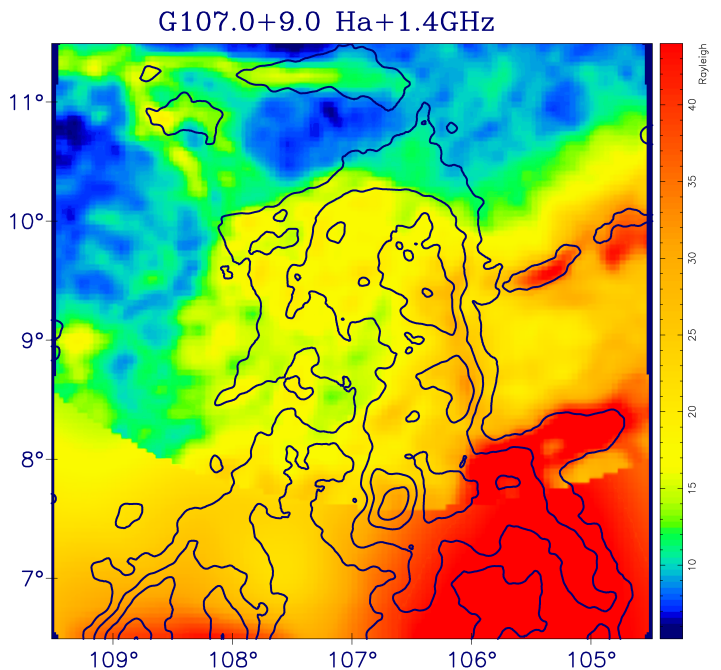}
\caption{Colour-coded H$\alpha$ map centred on G107.0+9.0 with overlaid total-intensity 1.4-GHz contours from the EMLS with compact
sources from the CGPS survey and large-scale emission subtracted (see text). Contours start at 100~mK $T_\mathrm{b}$ and are 100~mK $T_\mathrm{b}$ apart.}
\label{CGPS+EMLS}
\end{figure}

\citet{Fesen20} (their Fig.~12) noted faint 1420-MHz continuum emission from the northern-sky survey \citep{Reich82}
possibly associated 
with G107.0+9.0. The 1420-MHz survey was carried out with the Stockert 25-m telescope with 36$\arcmin$ angular resolution. 
G107.0+9.0 is clearly visible on the Effelsberg Medium Latitude Survey (EMLS) 1.4-GHz maps \citep{Uyaniker98, Uyaniker99, Reich04}.
This section of the EMLS is not yet published. 
The CGPS 1420-MHz Galactic plane survey extension \citep{Landecker10} also covers the area of G107.0+9.0
with an angular resolution of $1\arcmin \times 1\arcmin~\times$ cosec(dec). The interferometric data need the addition of
large-scale structures from single-dish surveys for reliably tracing extended structures like G107.0+9.0. For the high-latitude CGPS 
extension, these large-scale structures were provided by the 1420-MHz Stockert survey for total intensities, and for polarisation by 
data from the DRAO 26-m telescope \citep{Wolleben06}. 
Both single-dish surveys have comparable angular resolutions.
The EMLS data were included in the CGPS 1420-MHz data of the Galactic plane, but not in its high-latitude extension, because the EMLS
data were not available at that time. The missing emission from the EMLS increases the influence of baseline fluctuations from the 
low-resolution surveys, which lowers the quality of extended source images. In their Fig.~3, \citet{Landecker10} showed that 
the data from the Effelsberg 100-m telescope are essential for baselines between 9~m and
35~m, corresponding to emission structures between about 100$\arcmin$ and about 30$\arcmin$ in size, while the interferometric and low-resolution surveys 
as part of the CGPS have a low weight and thus do not contribute much when combining all data. 
We added the total-intensity and linear-polarisation data from this section of the EMLS to the CGPS data  
in the same way as that already described by \citet{Landecker10}.
The combined 1420-MHz data show structures down to about $1\arcmin$ in size on an absolute zero-level for both, total intensities and 
linear polarisation. However, we note numerous compact unresolved sources that mask the faint emission from G107.0+9.0. 
We separated compact sources and extended emission in the same way as described in Sect.~2.3 
with a filtering beam of 3$\arcmin$, which results in a map with small-scale structures up to 3$\arcmin$ and a larger-scale intensity map. 
The small-scale
structures originate from the long-spacing data provided by the DRAO interferometer. We did not find any filamentary structures in the small-scale 
map, just compact sources. We convolved the small-scale map to the Effelsberg beam and subtracted it from the EMLS map, which more clearly shows the 
emission from the SNR G107.0+9.0. We subtracted the large-scale emission
by a 'twisted' plane, so the edge areas of the map are close to zero, and show the result in Fig.~\ref{CGPS+EMLS}.

\begin{figure}
\centering
\includegraphics[angle=-90, width=0.50\textwidth]{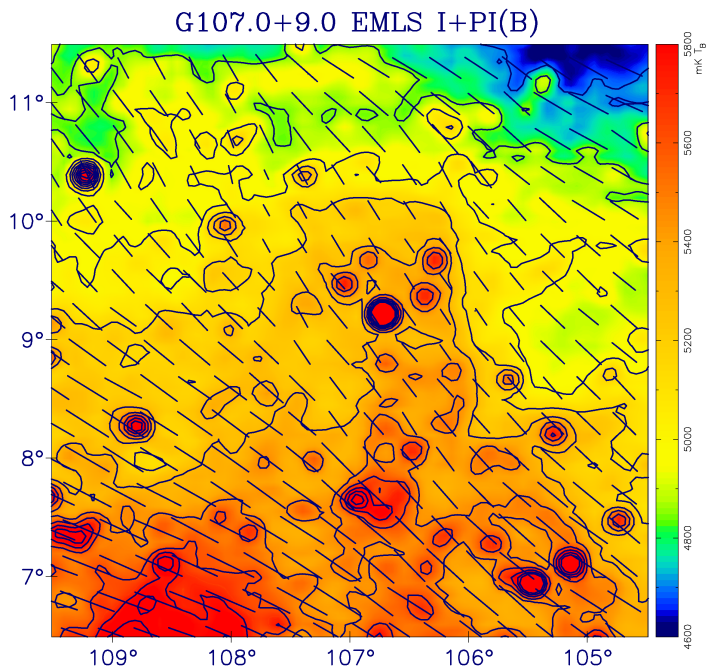}
\caption{Colour-coded EMLS 1.4-GHz total-intensity map of G107.0+9.0 at an absolute zero-level with overlaid polarised-intensity bars exceeding 
20~mK $T_\mathrm{b}$ in B-field direction. Total-intensity contours start at 
4.7~K~$T_\mathrm{b}$ and are 150~mK~$T_\mathrm{b}$ apart. }
\label{21cm}
\end{figure}

\begin{figure}
\centering
\includegraphics[angle=-90, width=0.50\textwidth]{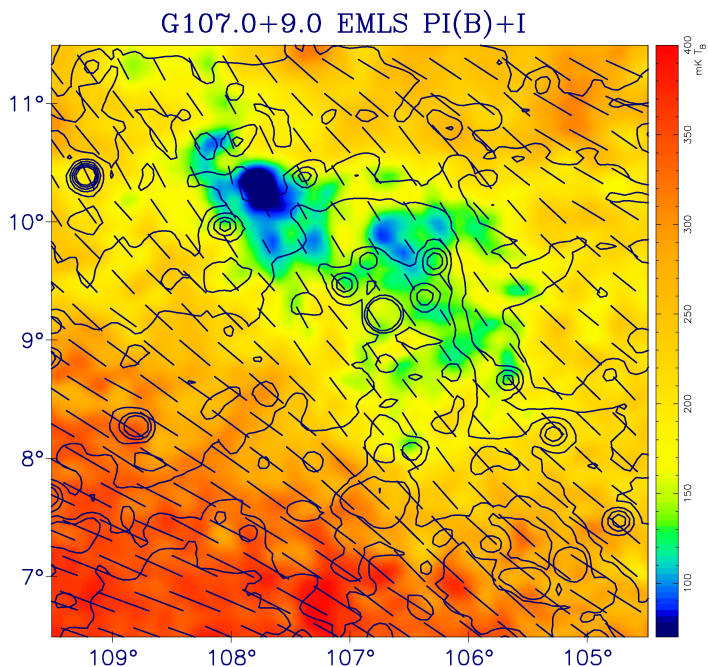}
\caption{Colour-coded EMLS 1.4-GHz absolutely calibrated polarised-intensity 
map of G107.0+9.0 with overlaid total-intensity contours starting at 4.7~K~$T_\mathrm{b}$ in steps of 150~mK~$T_\mathrm{b}$. 
Bars show $PA$s in B-field direction.}
\label{21cmPI}
\end{figure}

\begin{figure}
\centering
\includegraphics[angle=-90, width=0.50\textwidth]{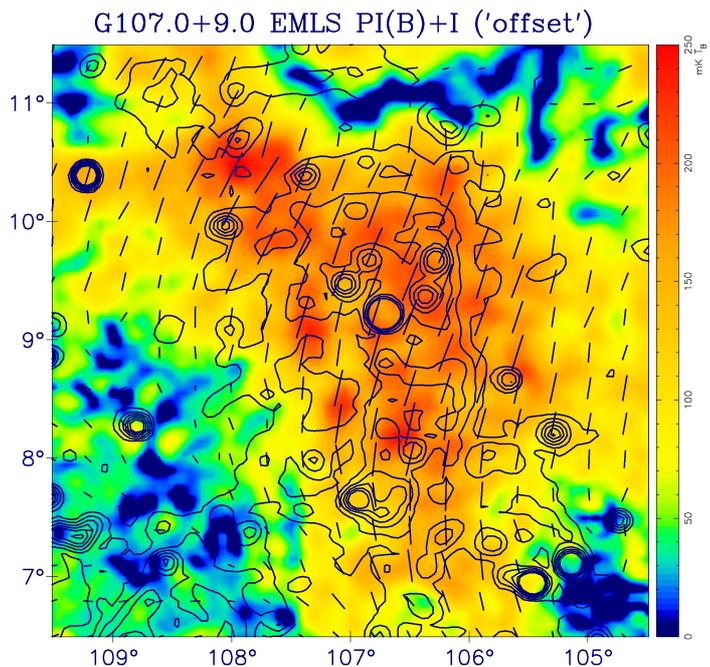}
\caption{Colour-coded EMLS 1.4-GHz image showing offset-subtracted polarised intensities of G107.0+9.0 with overlaid offset-corrected total-intensity contours 
starting at 0~mK~$T_\mathrm{b}$ in steps of 100~mK~$T_\mathrm{b}$. 
Polarised-intensity bars are in B-field direction.}
\label{21cmPItw}
\end{figure}

The 1.4-GHz total-intensity  EMLS map at an absolute zero level with overlaid bars of polarisation angles ${PA}$ is shown in Fig.~\ref{21cm}. The length of the bars 
is proportional to ${PI}$, which was calculated from the observed Stokes parameter $U$ and $Q$ as
$PI = \sqrt{U^2+Q^2-1.2\sigma_{U,Q}^{2}}$, including a polarisation zero-bias correction. $\sigma$ (=r.m.s.-noise)  is listed in 
Table~\ref{ObsTab} together with technical and calibration data for the EMLS. When adding 90$\degr$ to the polarisation angle $PA$ = 0.5 atan$(U/Q)$, 
the magnetic field direction (B-field) is indicated in the case of negligible Faraday rotation.  We show the EMLS ${PI}$ map in Fig.~\ref{21cmPI}.
Towards the centre of G107.0+9.0, ${PI}$ decreases by about 120~mK, while for a few small areas outside of G107.0+9.0, the decrease is up to  200~mK when compared 
to its surroundings. A depression of polarised emission is unexpected for an SNR, where polarised emission is usually excessive
compared to its surroundings. However, the latter case is valid for SNR maps, which typically are at a relative zero-level, where the edge areas of the observed $U$ and $Q$ 
maps were defined to be zero. The polarisation data from the EMLS are on an absolute level, so we subtracted a twisted plane from the
$U$ and $Q$ maps to obtain intensities near zero for the edge areas of the maps. The resulting $PI$~map is shown in Fig.~\ref{21cmPItw} with overlaid contours
of total intensities, which are also at a relative 
zero-level obtained in the same way as for $U$ and $Q$. We also show $PA$s, which deviate significantly in direction when compared with the $PA$s displayed in 
Fig.~\ref{21cmPI}. $PI$ is now stronger when compared to its surroundings and coincides with the enhanced total-intensity emission from G107.0+9.0, but also shows
enhanced $PI$ extending beyond the boundaries of the SNR, which is defined by its total-intensity emission. We estimated
$PI$ values in the range of 160~mK $T_\mathrm{b}$ to 170~mK $T_\mathrm{b}$ in the central area of G107.0+9.0, where total intensities
are around 220~mK $T_\mathrm{b}$. This corresponds to a percentage polarisation
of about 75\%, which is unusually high for SNRs and close to the upper limit for synchrotron emission. The polarisation maps including the high-resolution
interferometric CGPS data do not show small-scale polarised features from G107.0+9.0.

We measured an integrated 1.4-GHz flux density for G107.0+9.0 of 10.1 $\pm$ 1.5~Jy.

\begin{table*}[thp]
\caption{Observational parameters.}

\label{tab1}
\vspace{-1mm}
\centering
\begin{tabular}{lrrr}
\hline\hline
\multicolumn{1}{c}{} &\multicolumn{1}{c}{}  & \multicolumn{1}{c}{}\\
Data                           &Urumqi 4.8~GHz   &EMLS 1.4~GHz  \\
\hline             
Frequency [GHz]                                  &4.8                        &1.4 \\
Bandwidth [MHz]                                 &600                           &20   \\
HPBW[$\arcmin$]                                    &9.5                          &9.4 \\
Main calibrator                                      &3C286                        &3C286 \\
Flux density of 3C286 [Jy]                                 &7.5                         &14.4 \\
Polarisation percentage of 3C286 [\%]                       &11.3                      &9.3\\
Polarisation angle of 3C286 [$\degr$]                        &33                           &32 \\
r.m.s. ($I/U,Q$) [mK~$T_\mathrm{b}$]                &1.0/0.5                             &15/8 \\
\hline\hline
\end{tabular}
\label{ObsTab}
\vspace{-1mm}
\end{table*}

\begin{table*}[thp]
\caption{Integrated flux densities of G107.0+9.0.}

\label{tab2}
\vspace{-1mm}
\centering
\begin{tabular}{lrrrr}
\hline\hline
\multicolumn{1}{c}{} &\multicolumn{1}{c}{} &\multicolumn{1}{c}{} &\multicolumn{1}{c}{}  &\multicolumn{1}{c}{}\\            
Frequency [MHz]                 &22          &408          &1400      &4800     \\
\hline
Flux density  [Jy]                  &458         &23             &10.1         &2.6       \\
Flux density error [Jy]          &60            &3             &1.5           &0.4           \\
\hline\hline
\end{tabular}
\label{Flux}
\vspace{-1mm}
\end{table*}

\subsection{Urumqi medium-latitude 4.8-GHz data}

The Sino-German 4.8-GHz polarisation survey of the Galactic plane within latitudes of $\pm5\degr$ was carried out with the 
Urumqi 25-m telescope between 2004 and 2009 and published in a series of papers \citep {Gao10, Xiao11, Sun11a}. A review and 
summary of the 4.8-GHz survey results was given by \cite{Han15}. The survey has an angular 
resolution of $9\farcm5$ and a r.m.s.-sensitivity as listed in Table~\ref{ObsTab}, where also calibration data are listed. 
Observations of regions with higher Galactic latitudes 
were made 
at times when the Galactic plane was at too low elevations to allow high-quality mapping observations for the survey. The properties of the Urumqi survey, 
the observational set-up, and its reduction and calibration procedures were already described in detail in \citet{Sun07} and were applied to the 
medium-latitude regions as well. Diffuse polarised emission requires an absolute zero-level for its correct interpretation, which is not only
mandatory for interferometric data but also for single-dish maps 
\citep[e.g.][]{reich06}. These data were provided for the observed Urumqi polarisation
data by extrapolating WMAP absolute polarisation data at 22.8~GHz \citep{Hinshaw09}, which we also used for the high-latitude region assuming a typical 
temperature spectral index for Galactic synchrotron emission of $\beta$ = -3 with $T_\mathrm{b} \sim \nu^{+\beta}$. 

\begin{figure}
\centering
\includegraphics[angle=-90, width=0.50\textwidth]{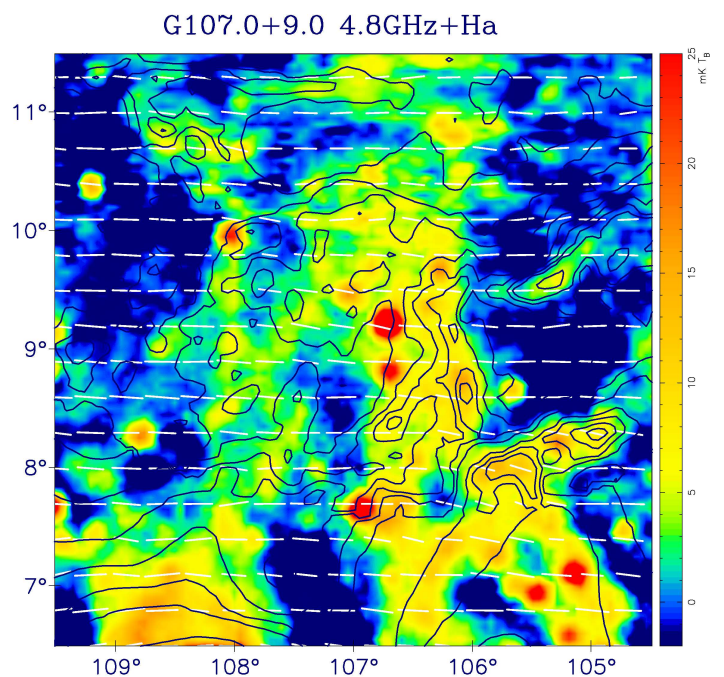}
\caption{Colour-coded 4.8-GHz total-intensity map of G107.0+9.0 with overlaid polarised-intensity bars along B-field direction.
Contour lines show H$\alpha$ intensities starting at 10~Rayleigh in steps of 3~Rayleigh.}
\label{6cm}
\end{figure}

We show the total-intensity map with superimposed polarisation bars in Fig.~\ref{6cm}. Extended total-intensity emission is
faint with some stronger compact background sources in the area of G107.0+9.0. After subtraction of compact sources from the field, we
obtained an integrated flux density of 2.6$\pm$0.4~Jy at 4.8~GHz. Polarisation is dominated by uniform large-scale Galactic
emission, where $PA$s show the B-field direction to run almost parallel to the
Galactic plane.

\begin{figure}
\centering
\includegraphics[angle=-90, width=0.50\textwidth]{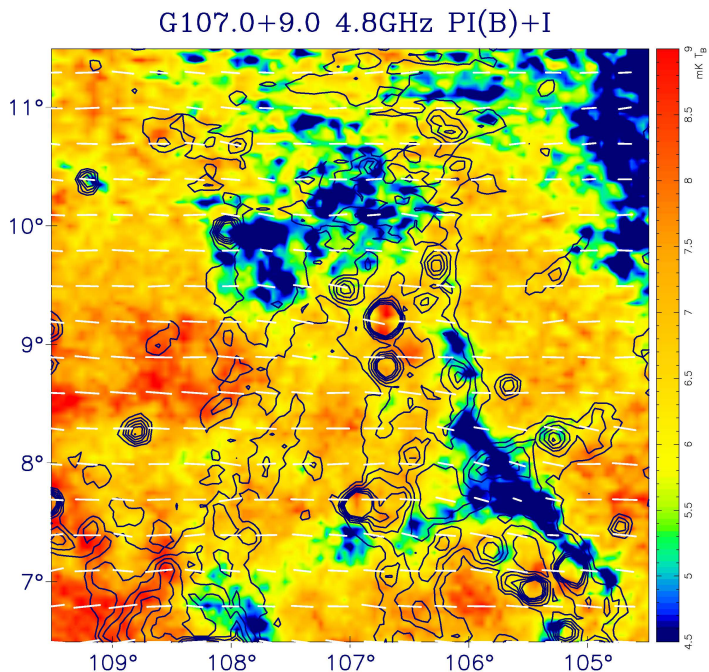}
\caption{Colour-coded 4.8-GHz polarised-intensity map of G107.0+9.0 with overlaid polarised-intensity bars in B-field direction.
Contour lines show total intensities starting at 3~mK~$T_\mathrm{b}$ 
in steps of 3~mK~$T_\mathrm{b}$.}
\label{6cmPI}
\end{figure}

\begin{figure}
\centering
\includegraphics[angle=-90, width=0.50\textwidth]{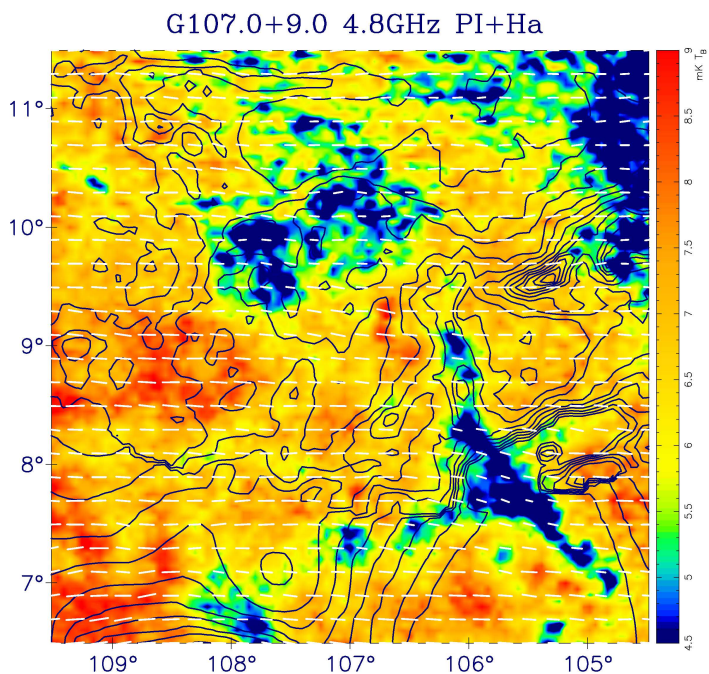}
\caption{Colour-coded 4.8-GHz absolute polarised-intensity map of G107.0+9.0 with overlaid polarised-intensity bars in the B-field direction.
Contour lines show H$\alpha$ intensities starting at 10~Rayleigh in steps of 3~Rayleigh.}
\label{6cmPIHa}
\end{figure}

\begin{figure}
\centering
\includegraphics[angle=-90, width=0.50\textwidth]{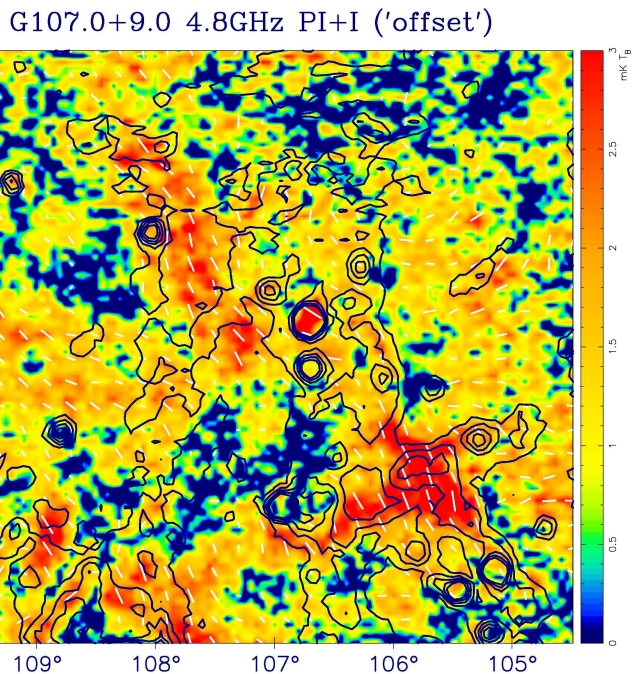}
\caption{Colour-coded 4.8-GHz polarised-intensity map at relative zero level with overlaid polarised-intensity bars shown in the B-field direction.
Contour lines show total intensities on a relative zero level starting at 
3~mK~$T_\mathrm{b}$ in steps of 4~mK~$T_\mathrm{b}$.}
\label{6cm-rel}
\end{figure}

Except for a  faint depolarised filament, 
G107.0+9.0 is not very clearly seen in polarised emission at 4.8~GHz (Fig.~\ref{6cmPI}) when compared to the EMLS map (Fig.~\ref{21cmPI}). 
Figure~\ref{6cmPIHa} shows the 4.8-GHz polarised
emission with overlaid H$\alpha$ intensity contours to indicate the relation between enhanced H$\alpha$ intensities and 
depolarisation.
This result remains unchanged after processing the 4.8-GHz polarisation data in a similar way to that done at 1.4~GHz, where $I, U$, and $Q$ are on 
a local zero level by subtracting a twisted plane defined by the edge-area intensities from the observed maps (Fig.~\ref{21cmPItw}). The result is shown in 
Fig.~\ref{6cm-rel}.
An extended broad polarisation ridge extends across the entire map along the magnetic field direction and is not limited to the 
area of G107.0+9.0. This is unexpected for an SNR, and we conclude that most of the polarised emission at 1.4~GHz 
and 4.8~GHz does not originate from SNR G107.0+9.0 itself, but likely from an associated Faraday screen (FS) acting on the polarised Galactic background
emission as discussed in Sect.~3.2.2.

\subsection{Flux densities and spectrum} 

\begin{figure}%
\centering
\includegraphics[angle=-90, width=0.50\textwidth]{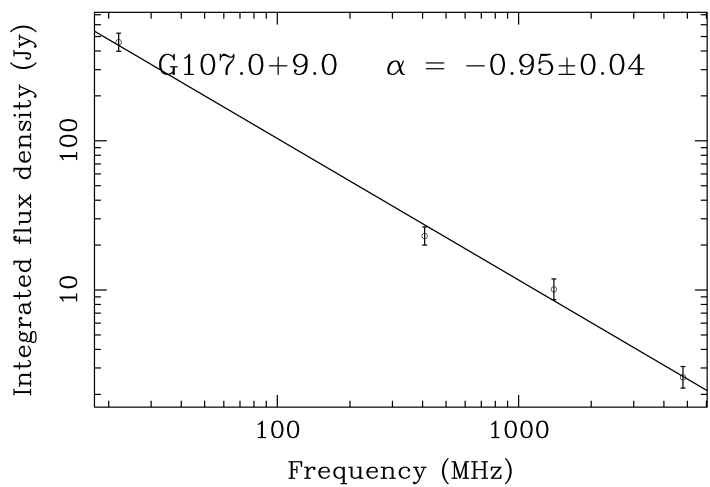}
\caption{Integrated flux-density spectrum of G107.0+9.0 including a least-square fit to the data.}
\label{spec}
\end{figure}

Table~\ref{Flux} lists the obtained integrated flux densities of G107.0+9.0, where we subtracted the flux densities of compact
sources before integrating in rings up to a radius of 1$\fdg$5 starting at the centre of G107.0+9.0. For the low-resolution 22~MHz map,
we performed a Gaussian fit and subtracted 10~Jy to account for the contribution of the strong source 4C +66.24 in the field. The 
contributions from fainter sources in the area of G107.0+9.0 could be ignored, because their contribution is about the same for the local zero level. 
Figure~\ref{spec} shows the resulting integrated flux-density spectrum of G107.0+9.0, which covers
a wide frequency range. The spectrum is fitted well and results in a spectral index of 
$\alpha = -0.95 \pm 0.04$ ($S \sim \nu^{+\alpha}$). 

\section{Discussion}

\subsection{Total intensities}

The total-intensity spectrum of G107.0+9.0 has a non-thermal spectral index of $\alpha = -0.95\pm0.04$, which is in agreement with its
identification as a non-thermal SNR, where its morphology resembles an asymmetric thick-shell SNR. The spectrum of G107.0+9.0 is steeper than 
that of the majority of shell-type SNRs in the adiabatic phase, which is around $\alpha = -0.5$ \citep{Dubner15} and may support the 
view of an SNR in the radiative phase of evolution \citep{Fesen20}. The surface brightness of G107.0+9.0 of 
$\sim5.4\times10^{-23}$~W m$^{-2}$Hz$^{-1}$sr$^{-1}$ at 1~GHz is among the lowest of the about 300 Galactic SNRs known at present. 
It is at the surface brightness level of G152.4-2.1 \citep{Foster13}, which was the lowest SNR surface brightness known in 2013. It is about
five times brighter than G181.1+9.5, a faint shell-type SNR outside of the Galactic plane, located in the Galactic anti-centre  
at the same latitude as G107.0+9.0. Currently, G181.1+9.5 is the SNR with the lowest known surface brightness \citep{Kothes17}. 

\subsection{Polarised emission}

The Effelsberg and Urumqi maps at 1.4~GHz and 4.8~GHz include linear-polarisation data. Detected polarised emission from 
G107.0+9.0 will strengthen its identification as an SNR. The Urumqi 4.8-GHz data do not show strong
polarised emission. 
A depolarised filament is seen along the western periphery of G107.0+9.0.  In contrast, we note
a clear polarised signal across the entire area of G107.0+9.0 at 1.4~GHz. We also note enhanced polarised emission extending beyond the boundary of G107.0+9.0, which is unexpected if it is related to the SNR.
We see a depression in polarised emission on the absolute zero-level EMLS map (Fig.~\ref{21cmPI}). When subtracting the surrounding polarised emission
in $U$ and $Q$,
polarised emission from G107.0+9.0 is visible (Fig.~\ref{21cmPItw}) with a rather different orientation of $PA$ compared to the absolute zero-level map.

This can be interpreted in two ways. Either the local magnetic field in the G107.0+9.0 area has a different orientation compared to the large-scale 
Galactic field, so the polarised emission we see towards G107.0+9.0 is reduced when added with the diffuse Galactic emission along the
line of sight; 
alternatively, G107.0+9.0 hosts a regular magnetic field component along the line of sight that acts as an FS. In this
case, the Galactic background emission beyond G107.0+9.0 is rotated and adds to the polarised foreground emission
of G107.0+9.0 causing a depression with respect to its undisturbed surroundings. 

At 1.4~GHz, we derived a percentage polarisation for G107.0+9.0 of about 75\%, but do not see corresponding polarised emission at a similar
percentage level at 4.8~GHz, which is unexpected because of decreasing depolarisation towards higher frequencies. 
Within the area of G107.0+9.0, the polarised signal is less uniform than at 1.4~GHz, showing emission patches of about 1~mK - 2~mK $T_\mathrm{b}$ in $PI$. 
This is below the expected level of about 5~mK when extrapolating 1.4-GHz $PI$ for a steep spectrum with a temperature spectral index $\beta = -3$. 
The maximum polarised signal in the $PI$-map at 4.8~GHz is about 3.7~mK $T_\mathrm{b}$; however, outside of G107.0+9.0
at $\it l,b$ = 105$\fdg7,8\fdg0$ (Fig.~\ref{6cm-rel}). The morphology of 
the diffuse 1.4-GHz polarised emission and the rather uniform distribution of $PA$s (see Fig.~\ref{21cmPItw}) are untypical for shell-type SNRs. 
In particular, it is unusual that the polarised emission partly extends beyond the total-intensity boundaries of G107.0+9.0. This
is most clearly seen towards the northeast of G107.0+9.0. From the polarisation properties at 1.4~GHz and 4.8~GHz, we conclude that 
the major portion of the polarised emission at 1.4~GHz results from an FS effect, most likely associated with G107.0+9.0 and its surroundings.  

\subsubsection{Rotation measures of extragalactic sources and pulsars in the area of G107.0+9.0}

Rotation measures ($RM$s) of polarised extragalactic sources in the G107.0+9.0 area were selected from the catalogue by \cite{Xu14}\footnote{http://zmtt.bao.ac.cn/RM/}.
$RM$ describes the
wavelength-dependent $PA_{\lambda}$, which deviates from
the intrinsic polarisation angle  $PA_{0}$,  via $PA_{\lambda}$ = $RM \times \lambda^{2}$ + $PA_{0.}$ From $RM$ = 0.81 $n_{\mathrm{e}}$[cm${^{-3}}$] $B{_{||}} [\mu$G] $L$[pc], we obtain information
about the regular magnetic component $B{_{||}}$ along the line of sight and the thermal electron density $n_{\mathrm{e}}$.
In general, the $RM$s of the extragalactic sources are almost all negative, with an
average $RM$ of about -51~rad~m$^{-2}$ for the area centred on G107.0+9.0 with a radius of 2$\degr$ to 3$\degr$.  $RM$s
decrease outside of the field at the same latitudes to about -37~rad~m$^{-2}$.  
\cite{Xu14} quote an r.m.s scatter for the mean $RM$s of about 9~rad~m$^{-2}$. The $RM$ data indicate an increase of the $RM$ caused 
by the FS of the order of -10~rad~m$^{-2}$ to -20~rad~m$^{-2}$. 

There is one pulsar with measured $RM$ in the field of G107.0+9.0.  PSR B2224+65 at $l,b$ = 108$\fdg638, 6\fdg$846, which is about $0\fdg4$ offset 
from the map centre, has an $RM$ of  -22.99~rad~m$^{-2}$ \citep{Noutsos15}. Its distance is  900~pc based on its dispersion measure.  
Thus, the pulsar is closer than G107.0+9.0 according to the 1.5~kpc to 2~kpc distance quoted by \citet{Fesen20}. Its smaller absolute $RM$ is in 
accordance with that of extragalactic sources,
which trace the magnetised interstellar medium throughout the entire Galactic disk, and indicate a magnetic field direction pointing
away from us throughout the Galaxy.

\subsubsection{Faraday-screen model for G107.0+9.0}

We apply the FS model by \citet{Sun07} to the 1.4-GHz polarisation data, which allows us to
calculate the properties of an FS along the line of sight and to separate foreground (fg) and background (bg) emission components. 
The observed polarised emission $PI_{\mathrm{on}}$ towards the FS (`on'-position) is compared with the emission outside $PI_{\mathrm{off}}$ 
(`off'-position). The difference of the polarisation angles $PA_{\mathrm{on}} - PA_{\mathrm{off}}$ is also required by the model.
Parameter $c$ is the ratio $PI_{\mathrm{fg}}$/$(PI_{\mathrm{fg}}+PI_{\mathrm{bg}})$  
and $\psi_{s}$ is the angle rotation caused by the FS. 
The parameter $f$ describes the depolarisation of the FS, where 1 stands for no and 0 for total depolarisation.
Using Eq.~(1), we obtained values for $c$ and  $\psi_{s}$, while $f$ has to be estimated:

\begin{equation}
\centering
\displaystyle{
\left\{
\begin{array}{cc}
\displaystyle
\frac{PI_{\mathrm{on}}}{PI_{\mathrm{off}}}=\sqrt{\mathit{f}^2(1-c)^2+c^2+2\mathit{f}c(1-c)\cos2\psi_s}\ , \\ \displaystyle
PA_{\mathrm{on}} - PA_{\mathrm{off}}=\frac{1}{2}\arctan\left(\frac{\mathit{f}(1-c)\sin2\psi_s}{c+\mathit{f}(1-c)\cos2\psi_s}\right).
 &
\end{array}
\right.
}
\label{eq1}
\end{equation}

$PA_{\mathrm{off}}$ varies between about
$140\degr$ and $160\degr$ in the surroundings of G107.0+9.0. We subtracted $150\degr$ from
the $PA$ map to  apply  Eq.~\ref{eq1}, which holds for $PA_{\mathrm{off}} = 0\degr$ 
\citep[see][]{Sun07}.
We measured  $PI_{\mathrm{on}}/PI_{\mathrm{off}} \sim 0.52$ with variations below 10$\%$
and $PA_{\mathrm{on}}-PA_{\mathrm{off}} \sim -15\degr$ with an uncertainty of about $\pm5\degr$. 

From Eq.~(1), we calculated an angle rotation by the FS of about $\psi_{s} \sim -65\degr$  ($f$ = 1).  At $\lambda$21\ cm (1.4~GHz), this corresponds to 
$RM \sim -25$~rad~m$^{-2}$.  We also calculated $c \sim 0.67$, which means that 67\%  of the observed polarised emission towards G107.0+9.0 
originates in the foreground of the FS.  Allowing some internal FS depolarisation,  $c$ decreases to 0.65/0.62 for $f$ = 0.9/0.8, while the corresponding
$RM$s are just marginally influenced and decrease to -24$\pm3$~rad~m$^{-2}$. Thus, moderate depolarisation within the FS
has little influence on the result from the model.
The $RM$s from the FS-model for G107.0+9.0 are in the range of the negative $RM$ excess of extragalactic sources towards G107.0+9.0 (see section above). This supports the
view that G107.0+9.0 and its immediate surroundings act as an FS. The effect of the FS on the 4.8-GHz polarisation is much smaller than at 1.4~GHz, because
a $RM$ of -25~rad~m$^{-2}$ causes an angle rotation of only about 6$\degr$ on the background polarisation. 
This agrees with the observations.

The offset-subtracted H$\alpha$ emission from G107.0+9.0  varies between 3~Rayleigh and 7~Rayleigh. An exception is the stronger H$\alpha$-filament at 
its western periphery with about 12~Rayleigh. For a distance of 1.5~kpc to 2~kpc, we find extinction values E(B-V) of around 0.4~mag  based on the E$_{g-r}$
values quoted by \citet{GGreen19}. Using Eq.~\ref{eq2}, we obtained extinction-corrected emission measures ($EM$)  between 20~pc cm$^{-6}$  and 50~pc cm$^{-6}$
for an electron temperature of 10,000~K ($T_{4}$). Electron temperatures of 4,000~K and 6,000~K reduce $EM$ by  a factor of 0.44 and 0.63, respectively.  We do
not know the electron temperature of the H$\alpha$ -emitting gas and assume an $EM$ of 25~pc cm$^{-6}$ in the following calculations: 

\begin{equation}
\centering
EM = 2.75~T_{4}^{0.9}~I_{\mathrm{H\alpha}}~exp[2.44E(B-V)].
\label{eq2}
\end{equation}

The size $L$ of G107.0+9.0 is between 75~pc and 100~pc according to \citet{Fesen20}.  From $EM = n_{\mathrm{e}}^2 \times L$, we calculated an average electron
density of $n_{\mathrm{e}}$ = 0.6 cm$^{-3}$ or 0.5 cm$^{-3}$.  Combining $n_{\mathrm{e}}$ with the derived $RM$ of -25~rad~m$^{-2}$ of the FS, we calculated the regular 
magnetic field component of the FS along the line of sight by $RM$ = 0.81 $n_{\mathrm{e}}$[cm${^{-3}}$] $B{_{||}} [\mu$G] $L$[pc]
to be 0.7~$\mu$G and 0.6~$\mu$G for $L$ = 75~pc and 100~pc, respectively:  

\begin{equation}
\centering
n_{e} = \sqrt{\frac{EM}{\mathit{f_\mathrm{{n_{e}}}}L}}~ {\rm cm^{-3}}.
\label{eq3}
\end{equation}

Thermal gas is not perfectly uniformly distributed, so a filling factor $f_\mathrm{{n_e}}$ has to be considered. The filling factor is below 1 and
increases the electron density  according to Eq.~(3). For instance, assuming $f_\mathrm{{n_e}}$ = 0.5, $n_{\mathrm{e}}$ raises to  0.8 cm$^{-3}$ 
or 0.7 cm$^{-3}$. 
A filling factor will also increase $B{_{||}}$ by $B{_{||}} / \sqrt{f_\mathrm{{n_e}}}$, which for  $f_\mathrm{{n_e}}$ = 0.5, increases  $B{_{||}}$
to 1.0~$\mu$G or 0.85~$\mu$G for $L$ = 75~pc or 100~pc, respectively.
Even for a small filling factor of $f_\mathrm{{n_e}}$ = 0.1, $B{_{||}}$ will be just about 2~$\mu$G. The regular field is stronger in the case
that it is inclined with respect to the line of sight. Anyway, $B$ is low for a compressed SNR magnetic field and of the order of the regular halo field according to 
the 3D model of \citet{Sun10}.  The positional coincidence of the FS with the SNR  suggests a close connection, while it remains unclear from the present
data how this magnetic field component may have been formed. If the magnetic field was preexisting, the SNR shock-accelerated electrons may have
escaped and radiated in the magnetic field.  

We made the same FS analysis for the depolarised filament at 4.8~GHz extending along the western periphery of G107.0+9.0. With all the uncertainties as we listed 
above, we estimated $EM$ to be about 
80~pc cm$^{-6}$. The intrinsic size of the filament is about 14$\arcmin$, which corresponds to a diameter of 6.5/8.7~pc for a 1.5/2~kpc distance. We calculated 
electron densities $n_{\mathrm{e}}$ between 3.0 cm$^{-3}$ and 3.5 cm$^{-3}$.
If this filament traced a part of the SNR shell seen edge-on, its extent along the 
line of sight will be larger and the electron density will decrease. Figure~\ref{6cmPIHa}, however, shows the depolarised filament to apparently continue outside of 
the SNR towards \it{l,b} \rm = 107$\degr$,7$\degr$, but this depolarised structure is unrelated to G107.0+9.0.    
For $PI_{\mathrm{on}}/PI_{\mathrm{off}} \sim 0.54$ and an angle difference $PA_{\mathrm{on}}-PA_{\mathrm{off}} \sim -7\degr$, the FS 
model results in a  $RM$ of about -300~rad~m$^{-2}$ for $f$ values of 0.6 and 0.7. The parameter $c$ is between 0.66 and 0.69 and agrees with the value we 
found for G107.0+9.0, what means that the filament is at about the same distance.  
We calculated a regular magnetic field $B{_{||}}$ along the line of 
sight of the order of 15~$\mu$G, which is significantly higher than the typical magnetic field in the interstellar medium out of the Galactic disc and requires compression.  


\subsubsection{Synchrotron emissivity towards G107.0+9.0}

From the FS model result, we found that about 67\% of the observed polarised emission towards G107.0+9.0 originates in its foreground, so we can
estimate the synchrotron emissivity. When the depolarisation does not increase with distance, we estimated the total-intensity component 
towards G107.0+9.0 by also assuming 67\% of the observed signal as a lower limit.  At 1.4~GHz, we measured typically about 4.9~K $T_\mathrm{b}$ at 9$\degr$ 
latitude at a few degrees in longitude `off' from G107.0+9.0 with an uncertainty of about 0.2~K. Correcting by 2.8~K $T_\mathrm{b}$ for the isotropic CMB and unresolved 
background components \citep{Reich88b}, we have 2.1~K $T_\mathrm{b}$ of Galactic emission. 67\% corresponds to an emissivity of 0.94~K $T_\mathrm{b}$/kpc and 
0.70~K $T_\mathrm{b}$/kpc for distances of G107.0+9.0 of 1.5~kpc and 2~kpc, respectively. These values are uncertain by about 10\%, but in any case at the 
upper end of quoted Galactic emissivities as discussed by \citet{Wolleben04} and agree with the view of a locally (<1~kpc) enhanced Galactic synchrotron 
emissivity \citep{Sun08,Sun10}. 

\section{Summary}

We extracted faint radio emission from the optically identified SNR G107.0+9.0 from published 
surveys at 22~MHz and 408~MHz and new observations at 1.4~GHz and 4.8~GHz. 
We found a non-thermal integrated spectral index of $\alpha = -0.95 \pm 0.04$. This spectrum is   
steeper than that of typical 
shell-type SNRs in the adiabatic evolution phase, which have a spectral index of about $\alpha \sim -0.5$. G107.0+9.0 does not show
the typical morphology of a shell-type SNR and may be in the radiative phase. Its surface brightness of 
$\sim 5.4\times10^{-23}$~W m$^{-2}$Hz$^{-1}$sr$^{-1}$  at 1~GHz is among the lowest currently known for SNRs. 

G107.0+9.0 is clearly seen in depolarisation at 1.4~GHz and also at 4.8~GHz. 
The polarised emission extends beyond the boundaries of G107.0+9.0, so that the 
polarised emission is interpreted as the result of an FS hosting a faint ordered magnetic field along the line of sight. 
The FS scenario fits all observational data, although with significant uncertainties.
The analysis of a depolarised filament at 4.8~GHz along the western periphery of G107.0+9.0 reveals a compressed regular magnetic 
field of about 15~$\mu$G.

\begin{acknowledgements}
This research is partly based on observations with the Effelsberg 100-m telescope of the MPIfR.
X.Y.~Gao is supported by the National Natural Science Foundation of China (Grant Nos.~11988101, U1831103).

\end{acknowledgements}

\bibliographystyle{aa}
\bibliography{bbfile}

\end{document}